\documentclass{ifacconf}

\usepackage{graphicx}      
\usepackage{natbib}        
\usepackage{amsmath,amssymb,amsfonts}
\usepackage{algorithmic}
\usepackage{graphicx}
\usepackage{textcomp}
\usepackage{xcolor}
\usepackage{tabularx}
\usepackage{diagbox}
\usepackage{slashbox}
\usepackage{multirow}
\usepackage{cancel}
\begin{document}
\begin{frontmatter}

\title{Analytical Gradient and Hessian Evaluation for System Identification using State-Parameter Transition Tensors} 


\author{Premjit Saha,} 
\author{Tarunraj Singh} 


\address{Mechanical and Aerospace Engineering, University at Buffalo, Buffalo, NY 14260, USA}

\begin{abstract}                

In this work, the Einstein notation is utilized to synthesize state and parameter transition matrices, by solving a set of ordinary differential equations. Additionally, for the system identification problem, it has been demonstrated that the gradient and Hessian of a cost function can be analytically constructed using the same matrix and tensor metrics. A general gradient-based optimization problem is then posed to identify unknown system parameters and unknown initial conditions. Here, the analytical gradient and Hessian of the cost function are derived using these state and parameter transition matrices. The more robust performance of the proposed method for identifying unknown system parameters and unknown initial conditions over an existing conventional quasi-Newton method-based system identification toolbox (available in MATLAB) is demonstrated by using two widely used benchmark datasets from real dynamic systems. In the existing toolbox, gradient and Hessian information, which are derived using a finite difference method, are more susceptible to numerical errors compared to the analytical approach presented.
\end{abstract}

\begin{keyword}
Gradient-based Optimization, Transition matrix and tensors, Gradient and Hessian, System identification.
\end{keyword}

\end{frontmatter}
\maketitle

\section{Introduction}

The primary goal of system identification is to estimate or infer models, typically in the form of mathematical equations or transfer functions, that capture the dynamics and relationships within a system. Often it involves evaluation of system parameters and initial conditions such that the states derived using the mathematical model for the system closely follow the observed states of the system with a desired degree of accuracy. 
There is a vast body of literature presenting algorithms and methods for system identification~\cite{Astrom.1971, Eykhoff.1968}. \cite{Nandi.2018} have presented an adjoint sensitivity-based approach to determine the gradients and Hessians of cost functions for system identification of dynamical systems. \cite{More.1982} reported that including the Hessian in Newton's method enables faster (quadratic) convergence compared to methods using only first-order information like the gradient. 
\cite{Majji.2008} have presented an analytical approach for developing an estimation framework (called the $J^{th}$ Moment Extended Kalman Filter (JMEKF)) which can be used for system identification in conjunction with estimating states. While in their work, they used state transition tensors to compute different orders of statistical moments in the extended Kalman filter framework, this article presents a generalized approach to providing the analytical gradient and Hessian of a cost function using the same state and parameter transition matrices, along with other higher-dimensional tensors. The stated matrix and higher dimensional tensors are evaluated by solving a set of ordinary differential equations (ODEs). The concepts of state and parameter transition matrix and other higher dimensional state and parameter transition tensors are inspired from \cite{turner2008high}. 
There are also challenges, such as computational cost and issues with ill-conditioned Hessians, that need to be addressed \cite{More.1982,Broyden.1970}. In practice, variations like quasi-Newton methods (e.g., BFGS) are often used to tackle these challenges while retaining some benefits of second-order optimization \cite{Buckley.1978,Buckley.1978b,Coleman.1983}. However, the accuracy of quasi-Newton methods can be affected by the numerical errors accumulated from first-order information derived by explicit finite difference-based methods \cite{Sod.1978,Smith.1985,Kunz.1993}. 
The analytical gradient and Hessian developed using the proposed methodology can facilitate computationally efficient and numerically stable execution of necessary calculations in each iteration of the optimization process for a general Newton method-based system identification approach for dynamical systems. The accuracy of the proposed system identification technique has been tested on widely used benchmark datasets associated with real dynamic systems \cite{unknown, Wigren.2013}. The proposed method's performance is also compared with an existing conventional quasi-Newton method-based system identification toolbox, where both first-order and second-order information are derived using the finite difference method.

\section{Framework for estimating State and Parameter Transition Matrices}
A general form of a nonlinear system is chosen, which is written in state space form and is sensitive to the perturbation of both initial values of the states ($\boldsymbol{\overrightarrow{x}(t_0)}:=\boldsymbol{\overrightarrow{x}_0}$) and system parameters ($\boldsymbol{\overrightarrow{p}}$) of the system as follows:
\begin{subequations}
    \begin{equation}
    \begin{split}
       & \dot{\boldsymbol{\overrightarrow{x}}} (t,\boldsymbol{\overrightarrow{x}}_0, \boldsymbol{\overrightarrow{p}}) ~=~\boldsymbol{\mathcal{\overrightarrow{F}}} (\boldsymbol{\overrightarrow{x}}(t,\boldsymbol{\overrightarrow{x}}_0, \overrightarrow{p})) \\
       & \text{where} \quad \boldsymbol{\overrightarrow{x}} (t,\boldsymbol{\overrightarrow{x}}_0, \overrightarrow{p})~, ~\boldsymbol{\mathcal{\overrightarrow{F}}} (\boldsymbol{\boldsymbol{\overrightarrow{x}}}(t,\boldsymbol{\overrightarrow{x}}_0, \overrightarrow{p})) \in \mathbb{R}^N
    \end{split} \label{eqn1a}
\end{equation}
\begin{equation}
    \begin{split}
        & \dot{\boldsymbol{\overrightarrow{p}}}~=~0 \quad \text{where} \quad \boldsymbol{\overrightarrow{p}}\in \mathbb{R}^M. \\
    \end{split}\label{eqn1b}
\end{equation}
\end{subequations}
\noindent The nonlinear equation can also be rewritten in Einstein's tensorial notation as:
\begin{subequations}
    \begin{equation}
    \begin{split}
       & \dot{x}_{i_1} =\mathcal{F}_{i_1} (x_k (t,\boldsymbol{\overrightarrow{x}}, \boldsymbol{\overrightarrow{p}}))~~\text{where}~~i_1,k = 1,2,3 \dots N 
    \end{split} \label{eqn2a}
\end{equation}
\begin{equation}
    \begin{split}
& \dot{p}_{i_2}~=~ 0 \quad \text{where} \quad  i_2~=~1,2,3 \dots M
    \end{split}\label{eqn2b}
\end{equation}
\end{subequations}
The practice of representing equations in both vector form and Einstein's tensorial notation will continue in the beginning part of the section to establish a better understanding of the process. The solution for Eq. (\ref{eqn1a}-\ref{eqn1b}) with the initial condition $\boldsymbol{\overrightarrow{x}}(t_0)$ and system parameters ($\boldsymbol{\overrightarrow{p}}$) is:
\begin{equation}
     \boldsymbol{\overrightarrow{x}} (t,\boldsymbol{\overrightarrow{x}_0}, \boldsymbol{\overrightarrow{p}}) =\boldsymbol{\overrightarrow{x}_0}
+\int_{t_0}^t~\boldsymbol{\mathcal{\overrightarrow{F}}} (\boldsymbol{\overrightarrow{x}}(\tau,\boldsymbol{\overrightarrow{x}_0}, \boldsymbol{\overrightarrow{p}})~d\tau
\label{eqn3}\end{equation}

\noindent
Hence, from  Eq.(\ref{eqn3}), the sensitivity of the solution $\boldsymbol{\overrightarrow{x}}$ $(t,\boldsymbol{\overrightarrow{x}_0}, \boldsymbol{\overrightarrow{p}})$ with respect to the initial conditions ($\boldsymbol{\overrightarrow{x}_0}$), which is also defined as state transition matrix, can be expanded as:
\begin{equation}
    \begin{split}
        & \dfrac{\partial\boldsymbol{\overrightarrow{x}} (t,\boldsymbol{\overrightarrow{x}_0},\boldsymbol{\overrightarrow{p}})}{\partial\boldsymbol{\overrightarrow{x}_0} } = \boldsymbol{\mathcal{I}}_{N \times N} \\
        & + \int_{t_0}^t\Bigg(\dfrac{\partial \boldsymbol{\mathcal{\overrightarrow{F}}}(\boldsymbol{\overrightarrow{x}})}{\partial\boldsymbol{\overrightarrow{x}}}\dfrac{\partial \boldsymbol{\overrightarrow{x}} (\tau,\boldsymbol{\overrightarrow{x}_0}, \boldsymbol{\overrightarrow{p}})}{\partial\boldsymbol{\overrightarrow{x}_0}} +\dfrac{\partial\boldsymbol{\mathcal{\overrightarrow{F}}}(\boldsymbol{\overrightarrow{x}})}{\partial\boldsymbol{\overrightarrow{p}}} \dfrac{\partial \boldsymbol{\overrightarrow{p}}}{\partial \boldsymbol{\overrightarrow{x}_0}} \Bigg) d \tau \\
        & \dfrac{\partial x_{i_1}}{\partial x_{0_{j_1}} } = \delta_{i_1 {j_1}}
        + \int_{t_0}^t \Bigg(\dfrac{\partial\mathcal{F}_{i_1}}{\partial x_m}\dfrac{\partial x_m}{\partial x_{0_{j_1}}} + \dfrac{\partial\mathcal{F}_{i_1}}{\partial p_n}\dfrac{\partial p_n}{\partial x_{0_{j_1}}} \Bigg) d\tau \\
        & \text{where}~i_1,m,j_1 = 1,2, 3  \dots  N~\text{and} ~ n= 1,2,3 \dots M.
    \end{split}\label{eqn4}
\end{equation}
\noindent
Eq.(\ref{eqn4}) can further be differentiated with respect to time to derive the following equation:
\begin{equation}
    \begin{split}
        & \dfrac{d}{dt}\left(\dfrac{\partial x_{i_1} (t,\boldsymbol{\overrightarrow{x}_0}, \boldsymbol{\overrightarrow{p}})}{\partial x_{0_{j_1}}} \right)=\dfrac{\partial\mathcal{F}_{i_1}}{\partial x_m} \dfrac{\partial x_m}{\partial x_{0_{j_1}}} + \dfrac{\partial\mathcal{F}_{i_1}}{\partial p_n}\dfrac{\partial p_n}{\partial x_{0_{j_1}}} \\
        & \text{with}~\dfrac{\partial x_{i_1} (t,\boldsymbol{\overrightarrow{x}_0}, \boldsymbol{\overrightarrow{p}})}{\partial x_{0_{j_1}}} \Bigg |_{t=t_0}=\delta_{i_1 {j_1}}: \text{Kronecker delta } \\
        & \text{or,}~ \dfrac{d\Phi_{{i_1},{j_1}}(t,t_0)}{dt} = \mathcal{F}_{x{i_1},m} (t) \Phi_{m,{j_1}}(t,t_0) + \mathcal{F}_{p~i_1,n} (t) \dfrac{\partial~p_n}{\partial x_{0_{j_1}}} \\
        & \text{with}~\dfrac{\partial x_{i_1} (t,\boldsymbol{\overrightarrow{x}_0},\boldsymbol{\overrightarrow{p}})}{\partial x_{0_{j_1}}}\equiv\Phi_{{i_1},{j_1}}(t,t_0)~,~\Phi_{i_1,{j_1}}(t_0,t_0)=\delta_{i_1 j_1}
    \end{split}\label{eqn5}
\end{equation}
\noindent
where $()_{x} = \dfrac{\partial ()}{\partial \boldsymbol{\overrightarrow{x}}}$ and $()_{p} = \dfrac{\partial ()}{\partial \boldsymbol{\overrightarrow{p}}}$. Similarly the sensitivity of the solution $\boldsymbol{\overrightarrow{x}} (t,\boldsymbol{\overrightarrow{x}_0}, \boldsymbol{\overrightarrow{p}})$ with respect to the set of system parameters $\boldsymbol{\overrightarrow{p}}$, which is also defined as parameter transition matrix, leads to the following equation:
\begin{equation}
    \begin{split}
                &\dfrac{\partial\boldsymbol{\overrightarrow{x}} (t,\boldsymbol{\overrightarrow{x}_0}, \boldsymbol{\overrightarrow{p}})}{\partial\boldsymbol{\overrightarrow{p}} } \\
                &= \dfrac{\partial\boldsymbol{\overrightarrow{x}_0}}{\partial\boldsymbol{\overrightarrow{p}}}+\int_{t_0}^t \Bigg(\dfrac{\partial\boldsymbol{\mathcal{\overrightarrow{F}}}(\boldsymbol{\overrightarrow{x}}(\tau,\boldsymbol{\overrightarrow{x}}_0,\overrightarrow{p}))}{\partial\boldsymbol{\overrightarrow{x}}}\dfrac{\partial\boldsymbol{\overrightarrow{x}} (\tau,\boldsymbol{\overrightarrow{x}_0}, \boldsymbol{\overrightarrow{p}})}{\partial\boldsymbol{\overrightarrow{p}}} \\ &+\dfrac{\partial\boldsymbol{\mathcal{\overrightarrow{F}}}(\boldsymbol{\overrightarrow{x}}(t,\boldsymbol{\overrightarrow{x}}_0, \overrightarrow{p}))}{\partial\boldsymbol{\overrightarrow{p}}}\Bigg) d \tau \\
               & \dfrac{\partial x_{i_1} (t,\boldsymbol{\overrightarrow{x}_0}, \boldsymbol{\overrightarrow{p}})}{\partial p_{j_2} } = \dfrac{\partial x_{0_{i_1}}}{\partial~p_{j_2}} + \int_{t_0}^t\left(\dfrac{\partial \mathcal{F}_{i_1}}{\partial x_m} \dfrac{\partial x_m}{\partial p_{j_2}}+\dfrac{\partial \mathcal{F}_{i_1}}{\partial p_{j_2}} \right) d \tau \quad \\
        & \text{where} ~i_1,m=1,2,3 \dots N ~ \text{and}~j_2=1,2,3 \dots M.
    \end{split}\label{eqn6}
\end{equation}
\noindent
It is further assumed that the initial conditions of the states ($\boldsymbol{\overrightarrow{x}}(t_0):=\boldsymbol{\overrightarrow{x}_0}$) and system parameters ($\boldsymbol{\overrightarrow{p}}$) of the system are not correlated i.e. $\dfrac{\partial\boldsymbol{\overrightarrow{x}} (t,\boldsymbol{\overrightarrow{x}_0}, \boldsymbol{\overrightarrow{p}})}{\partial\boldsymbol{\overrightarrow{p}}} \Bigg |_{t=t_0}=\dfrac{\partial~\boldsymbol{\overrightarrow{x}_0}}{\partial\boldsymbol{\overrightarrow{p}}}=\boldsymbol{0}_{N \times M}$ and $\dfrac{\partial\boldsymbol{\overrightarrow{p}}}{\partial\boldsymbol{\overrightarrow{x}} (t,\boldsymbol{\overrightarrow{x}_0}, \boldsymbol{\overrightarrow{p}})} \Bigg |_{t=t_0}=\dfrac{\partial\boldsymbol{\overrightarrow{p}}}{\partial\boldsymbol{\overrightarrow{x}_0}}=\boldsymbol{0}_{M \times N}$. Hence, Eq.(\ref{eqn5}-\ref{eqn6}) can be rewritten as
\begin{equation}
    \begin{split}
        & \dfrac{d \Phi_{{i_1},{j_1}} (t,t_0)}{dt}= \mathcal{F}_{x~{i_1},m} (t)\Phi_{m,{j_1}}(t,t_0)+\mathcal{F}_{p~i_1,n} (t)~\cancelto{0}{\dfrac{\partial~p_n}{\partial x_{0_{j_1}}}};\\
        & \text{or,}~ \dfrac{d\Phi_{{i_1},{j_1}} (t,t_0)}{dt}=\mathcal{F}_{x~{i_1},m} (t)\Phi_{m,{j_1}}(t,t_0); \\
        &\text{with} \quad  \Phi_{i_1,{j_1}}(t_0,t_0)=\delta_{i_1 j_1} \\
    \end{split} \label{eqn7}
\end{equation} 
\noindent
and
\begin{equation}
    \begin{split}
                & \dfrac{\partial x_{i_1} (t,\boldsymbol{\overrightarrow{x}_0}, \boldsymbol{\overrightarrow{p}})}{\partial p_{j_2}}=\cancelto{0}{\dfrac{\partial x_{0_{i_1}}}{\partial p_{j_2}}}+\int_{t_0}^t \Bigg(\dfrac{\partial \mathcal{F}_{i_1}}{\partial x_m} \dfrac{\partial x_m}{\partial p_{j_2}}+\dfrac{\partial\mathcal{F}_{i_1}}{\partial p_{j_2}} \Bigg)d\tau
    \end{split}\label{eqn8}
\end{equation}
\noindent
Eq.(\ref{eqn8}) can be differentiated with respect to time to derive the following equation:

\begin{equation}
    \begin{split}
    & \dfrac{d}{dt} \left(\dfrac{\partial x_{i_1} (t,\boldsymbol{\overrightarrow{x}_0}, \boldsymbol{\overrightarrow{p}})}{\partial p_{j_2}} \right)=\dfrac{\partial\mathcal{F}_{i_1}}{\partial x_m}~\dfrac{\partial x_m}{\partial p_{j_2}}+\dfrac{\partial \mathcal{F}_{i_1}}{\partial p_{j_2}} \\
        & \text{or,}~\dfrac{d\Theta_{{i_1},{j_2}}(t,t_0)}{dt}=\mathcal{F}_{x~{i_1},m} (t)\Theta_{m,{j_2}}(t,t_0)+\mathcal{F}_{p~i_1,j_2} (t); \\
        &\text{with}~\dfrac{\partial x_{i_1} (t,\boldsymbol{\overrightarrow{x}_0},\boldsymbol{\overrightarrow{p}})}{\partial p_{j_2}}\equiv\Theta_{{i_1},{j_2}}(t,t_0)~,~\Theta_{i_1,j_2}(t_0,t_0)=0_{i_1 j_2}
    \end{split}\label{eqn9}
\end{equation}
\noindent
Eq.(\ref{eqn1b}) can also similarly be used to result in the following equations:
\begin{subequations}
    \begin{equation}
    \dfrac{d}{d t}\left(\dfrac{\partial p_{i_2}}{\partial x_{j_1}}\right)=0_{i_2 j_1} \quad \text{with} \quad \dfrac{\partial p_{i_2}}{\partial x_{j_1}}\Bigg |_{t=t_0}=0_{i_2 j_1} \label{eqn10a}
\end{equation}
\begin{equation}
    \dfrac{d}{d t}\left(\dfrac{\partial p_{i_2}}{\partial p_{j_2}} \right)=0_{i_2 j_2} \quad \text{with} \quad \dfrac{\partial p_{i_2}}{\partial p_{j_2}}\Bigg |_{t=t_0}=\delta_{i_2 j_2}. \label{eqn10b}
\end{equation}
\end{subequations}
\noindent
Eq.(\ref{eqn10a}-\ref{eqn10b}) along with their initial conditions, ensure that the state transition tensors satisfy, $\dfrac{\partial\boldsymbol{\overrightarrow{p}}}{\partial\boldsymbol{\overrightarrow{x}_0}}=$ $\boldsymbol{0}_{M \times N}$ and $\dfrac{\partial \boldsymbol{\overrightarrow{p}}}{\partial \boldsymbol{\overrightarrow{p}}}=$ $\boldsymbol{\mathcal{I}}_{M \times M}$ for all time respectively. Similarly, the following higher dimensional transition tensor is defined
\begin{equation}
     \begin{split}
        &\dfrac{\partial}{\partial\boldsymbol{\overrightarrow{x}_0}}\left(\dfrac{\partial\boldsymbol{\overrightarrow{x}} (t,\boldsymbol{\overrightarrow{x}_0}, \boldsymbol{\overrightarrow{p}})}{\partial\boldsymbol{\overrightarrow{x}_0} }\right) = \\ &\int_{t_0}^t\Bigg(\left(\dfrac{\partial\boldsymbol{\overrightarrow{x}} (\tau,\boldsymbol{\overrightarrow{x}_0},\boldsymbol{\overrightarrow{p}})}{\partial\boldsymbol{\overrightarrow{x}_0}}\right)^T\dfrac{\partial}{\partial\boldsymbol{\overrightarrow{x}}}\left(\dfrac{\partial\boldsymbol{\mathcal{\overrightarrow{F}}}}{\partial\boldsymbol{\overrightarrow{x}}}\right)\dfrac{\partial\boldsymbol{\overrightarrow{x}} (\tau,\boldsymbol{\overrightarrow{x}_0},\boldsymbol{\overrightarrow{p}})}{\partial\boldsymbol{\overrightarrow{x}_0}} \\
        &+\dfrac{\partial\boldsymbol{\mathcal{\overrightarrow{F}}}}{\partial\boldsymbol{\overrightarrow{x}}}\dfrac{\partial}{\partial\boldsymbol{\overrightarrow{x}_0}}\left(\dfrac{\partial\boldsymbol{\overrightarrow{x}}(\tau,\boldsymbol{\overrightarrow{x}_0}, \boldsymbol{\overrightarrow{p}})}{\partial\boldsymbol{\overrightarrow{x}_0} }\right) \\
        &+\left(\dfrac{\partial\boldsymbol{\overrightarrow{x}}(\tau,\boldsymbol{\overrightarrow{x}_0}, \boldsymbol{\overrightarrow{p}})}{\partial\boldsymbol{\overrightarrow{x}_0}}\right)^T\dfrac{\partial}{\partial\boldsymbol{\overrightarrow{p}}}\left(\dfrac{\partial\boldsymbol{\mathcal{\overrightarrow{F}}}}{\partial\boldsymbol{\overrightarrow{x}}}\right)\dfrac{\partial\boldsymbol{\overrightarrow{p}}}{\partial\boldsymbol{\overrightarrow{x}_0}} \\
        &+\dfrac{\partial\boldsymbol{\mathcal{\overrightarrow{F}}}}{\partial\boldsymbol{\overrightarrow{x}}}\dfrac{\partial}{\partial\boldsymbol{\overrightarrow{p}}}\left(\dfrac{\partial\boldsymbol{\overrightarrow{x}}(\tau,\boldsymbol{\overrightarrow{x}_0}, \boldsymbol{\overrightarrow{p}})}{\partial\boldsymbol{\overrightarrow{x}_0} }\right)\dfrac{\partial\boldsymbol{\overrightarrow{p}}}{\partial\boldsymbol{\overrightarrow{x}_0}}\Bigg)d\tau
    \end{split}\label{eqn11}
\end{equation}
\noindent
Eq.(\ref{eqn11}) can be differentiated with respect to time, with the conditions depicted in  Eq.(\ref{eqn10a}) applied to derive the following equation: 
\begin{equation}
    \begin{split}
        &\dfrac{d \Phi^1_{i_1 j_1 k_1} (t,t_0)}{dt}=\dfrac{d \Phi_{i_1,j_1 k_1}(t,t_0)}{dt} \\
        &=\mathcal{F}_{xx~i_1,mn} (t)\Phi_{m,j_1}(t,t_0)\Phi_{n,k_1}(t,t_0) \\
        &+\mathcal{F}_{x~i_1,m} (t)\Phi_{m,j_1 k_1} (t,t_0) \\
        &\text{where} ~\Phi_{i_1,j_1 k_1} (t_0,t_0)\equiv\Phi^1_{i_1 j_1 k_1} (t_0,t_0)=\mathcal{O}, \\
        & \text{and}~~i_1,k_1,j_1, m, n=1,2,3 \dots N.
    \end{split}\label{eqn13}
\end{equation}

and $\dfrac{\partial^2 x_{i_1} (t,\boldsymbol{\overrightarrow{x}_0},\boldsymbol{\overrightarrow{p}})}{\partial x_{0_{k_1}} \partial x_{0_{j_1}} }\equiv\Phi_{i_1,j_1 k_1} (t,t_0)\equiv\Phi^1_{i_1 j_1 k_1} (t,t_0)$. Similarly, time evolution for the following higher dimensional transition tensor can be written as
\begin{equation}
    \begin{split}
        &\dfrac{d \Theta^1_{i_1 j_2 k_2}}{dt}=\dfrac{d \Theta_{i_1, j_2 k_2}}{dt}=\mathcal{F}_{xx~i_1,m n}\Theta_{m, j_2}\Theta_{n, k_2}\\   
        &+\mathcal{F}_{ p x~i_1 , j_2 m}\Theta_{m, k_2} +\mathcal{F}_{ x p~i_1 , m k_2}\Theta_{m, j_2} \\
        &+\mathcal{F}_{ x~i_1 , m}\Theta_{m, j_2 k_2}+\mathcal{F}_{ p p~i_1, j_2 k_2} \\
        & \text{with}~\Theta_{i_1,j_2 k_2} (t_0,t_0)\equiv~\Theta^1_{i_1 j_2 k_2} (t_0,t_0)=\mathcal{O}; \\
        & i_1,m,n=1,2,3 \dots N; \quad j_2,k_2=1,2,3 \dots M;
    \end{split}\label{eqn14}
\end{equation}

while
$\dfrac{\partial^2 x_i(t,\boldsymbol{\overrightarrow{x}_0},\boldsymbol{\overrightarrow{p}})}{\partial x_{k_2} \partial x_{j_2}}$$\equiv \Theta^1_{i_1 j_2 k_2}(t,t_0)$$\equiv \Theta_{i_1, j_2 k_2}(t,t_0) $.

Time evolution for two more higher dimensional transition tensors can be further developed. First,
\begin{equation}
    \begin{split}
& \dfrac{d\chi^1_{i_1 j_1 k_2}}{dt}=\dfrac{d}{dt}\left(\dfrac{\partial\Phi_{i_1,j_1}}{\partial p_{k_2}}\right) = \mathcal{F}_{xx~i_1,m n}\Phi_{m, j_1}\Theta_{n, k_2}\\
&+\mathcal{F}_{xp~i_1,m k_2}\Phi_{m,j_1} +\mathcal{F}_{x~i_1,m }\chi_{m j_1 k_2} \\
& \text{with}~\chi^1_{i_1 j_1 k_2} (t_0,t_0)=\mathcal{O};~ i_1,j_1,m,n=1,2,3 \dots N; \\
        &\text{and}~k_2=1,2,3 \dots M;
    \end{split}\label{eqn15}
\end{equation}

while
$\textstyle \dfrac{\partial^2 x_i (t,\boldsymbol{\overrightarrow{x}_0},\boldsymbol{\overrightarrow{p}})}{\partial p_{{k_2}}\partial x_{0_{j_1}}}\equiv\chi^1_{i_1 j_1 k_2} (t,t_0)$. Second,
\begin{equation}
    \begin{split}
        & \dfrac{d\chi^2_{i_1 j_2 k_1}}{dt}=\dfrac{d}{dt}\left( \dfrac{\partial\Theta_{i_1,j_2}}{\partial x_{0_{k_1}}}\right)=\mathcal{F}_{xx~i_1,m n}\Theta_{m, j_2}\Phi_{n, k_1}\\
        &+\mathcal{F}_{x~i_1,m }\chi_{m j_2 k_1}+\mathcal{F}_{ p x~i_1 , j_2 m}\Phi_{m, k_1} \\
        &\text{with}~\chi^2_{i_1 j_2 k_1} (t_0,t_0)=\mathcal{O};~i_1,k_1,m,n=1,2,3 \dots N; \\
        &\text{and}~j_2=1,2,3 \dots M; 
    \end{split}\label{eqn16}
\end{equation}

\noindent
while $\dfrac{\partial^2 x_i (t,\boldsymbol{\overrightarrow{x}_0}, \boldsymbol{\overrightarrow{p}})}{ \partial x_{0_{k_1}}\partial p_{{j_2}}} \equiv \chi^2_{i_1 j_2 k_1} (t,t_0) $

In this section, along with the governing differential equation in Eq.(\ref{eqn1a}-\ref{eqn1b}), a set of other differential equations have been developed depicting the evolution of state transition tensor $\boldsymbol{\Phi}$, parameter transition tensor $\boldsymbol{\Theta}$, and more higher dimensional state transition tensors ($\boldsymbol{\Phi^1}$, $\boldsymbol{\Theta^1}$, $\boldsymbol{\chi^1}$, and $\boldsymbol{\chi^2}$) with respect to time. 

\section{Gradient and Hessian using higher order transition tensors}

The cost ($J$) of the minimization problem is:
\begin{equation}
\begin{split}
      \text{min}~  J=&\sum_{h=0}^P \big(\boldsymbol{\overrightarrow{y}_{{obs}}}(t_h) -\boldsymbol{\overrightarrow{\tilde{y}}} (t_h)\big)^T \big(\boldsymbol{\overrightarrow{y}_{{obs}}}(t_h) -\boldsymbol{\overrightarrow{\tilde{y}}}(t_h) \big) \\
        \text{such that} &  \\
        \dot{x}_i (t_h)=&\mathcal{F}_i (x_k (t_h,\boldsymbol{\overrightarrow{x}_0},\boldsymbol{\overrightarrow{p}})) \\
        \tilde{y}_{g}(t_h)= & C_{g}\left(\boldsymbol{\overrightarrow{x}} (t_h,\boldsymbol{\overrightarrow{x}_0},\boldsymbol{\overrightarrow{p}}),\boldsymbol{\overrightarrow{p}}\right), ~ \text{where}~ C_g: \boldsymbol{\overrightarrow{x}} \to \tilde{y}_{g}\\
        \text{with}~i =&1,2,3 \dots N;~\text{and}~g=1,2,3 \dots S; 
\end{split}\label{eqn17}
\end{equation}
\noindent
In Eq.(\ref{eqn17}) $()_{obs}$ signifies measured variables, whereas $\tilde{()}$ signifies simulated variables from the system model. In this section, the state transition matrix and other higher dimensional state transition tensors will be used to construct the gradient and Hessian of the cost function. The gradient of the cost ($J$) can be defined as:
\begin{equation}
\begin{split}
&\dfrac{\partial J}{\partial x_{0_{j_1}}}=-2\sum_{h=0}^P\left( y_{obs_{g}}(t_h)-\tilde{y}_{g}(t_h)\right)~C_{x g,i }\Phi_{i,j_1}(t_h,t_0) \\
& \text{where} \quad i,j_1=1,2,3 \dots N; ~g = 1,2,3, \dots S; 
    \end{split}\label{eqn18a}
\end{equation}
and
\begin{equation}
    \begin{split}
        &\dfrac{\partial J}{\partial p_{j_2}}=-2\sum_{h=0}^P( y_{obs_{g}}(t_h)-\tilde{y}_{g}(t_h))\left(C_{x g,i }\Theta_{i, j_2}+C_{p g,j_2 }\right) \\
        & \text{where}~i=1,2,3 \dots N;~j_2=1,2,3 \dots M;\\
        & \text{and} ~g= 1,2,3, \dots S.
    \end{split}\label{eqn18b}
\end{equation}
\noindent
Similarly, the Hessian of the cost ($J$) can be defined as the following:
    \begin{equation}
        \begin{split}
        & \dfrac{\partial}{\partial x_{0_{k_1}}}\left(\dfrac{\partial J}{\partial x_{0_{j_1}}}\right) =-2\sum_{h=0}^P\left( y_{obs_{g}}(t_h) - \tilde{y}_{g}\right)\Big(C_{x~g,i }\Phi_{i, j_1 k_1} \\
        &+C_{xx~g,i \,m }\Phi_{i, j_1}\Phi_{m, k_1}\Big) +2\sum_{h=0}^P C_{x~g,m }\Phi_{m, k_1}~C_{x~g,i }\Phi_{i, j_1} \\
        & \text{where}~~i,j_1,k_1,m=1,2,3 \dots N;~\text{and} ~g= 1,2,3, \dots S;
        \end{split}
    \end{equation}
    \begin{equation}
        \begin{split}
            & \dfrac{\partial}{\partial p_{k_2}}\Big(\dfrac{\partial J}{\partial x_{0_{j_1}}}\Big)=\sum_{h=0}^P-2( y_{obs_{g}}(t_h) - \tilde{y}_{g}(t_h)) \Big(C_{x~g,i }\chi^1_{i j_1 k_2} \\
        & +C_{x p~g,i k_2 }\Phi_{i, j_1}+C_{x x~g,i \,m }\Phi_{i, j_1}\Theta_{m, k_2}\Big) \\
        &+2\sum_{h=0}^P\Big(C_{x~g,m }\Theta_{m, k_2}C_{x~g,i }\Phi_{i, j_1}+C_{p~g,k_2 }C_{x~g,i }\Phi_{i, j_1} \Big)\\
        & \text{where}~~i,j_1,m=1,2,3 \dots N;~k_2=1,2,3 \dots M; \\
        & \text{and} ~g= 1,2,3, \dots S;
        \end{split}
    \end{equation}

\begin{equation}
    \begin{split}
        & \dfrac{\partial}{\partial x_{0_{k_1}}}\Big(\dfrac{\partial J}{\partial p_{j_2}}\Big) = -2\sum_{h=0}^P( y_{obs_{g}}(t_h) - \tilde{y}_{g}(t_h))\Big(C_{x~g,i } \chi^2_{i j_2 k_1} \\
        &+C_{x x~g,i m }\Phi_{m, k_1}\Theta_{i,j_2}+C_{p x~g,j_2 m }\Phi_{m, k_1}\Big)\\
        & +2\sum_{h=0}^P\Big(C_{x~g,m }\Phi_{m,k_1}C_{x~g,i }\Theta_{i, j_2}+C_{x~g,m }\Phi_{m,k_1}C_{p~g,j_2 } \Big) \\
         & \text{where}~~i,k_1,m=1,2,3 \dots N;~j_2=1,2,3 \dots M; \\
        & \quad \quad ~~ ~g= 1,2,3, \dots S;
    \end{split}
\end{equation}
\noindent
and finally,
\begin{equation}
    \begin{split}
                & \dfrac{\partial}{\partial p_{k_2}}\left(\dfrac{\partial J}{\partial p_{j_2}}\right) \\
                &=-2 \sum_{h=0}^P\left( y_{obs_{g}}(t_h)-\tilde{y}_{g}(t_h)\right)\Big( C_{xx~g,i m }\Theta_{m,k_2}\Theta_{i,j_2} \\
                & +C_{x p~g,i k_2 }\Theta_{i,j_2}+C_{x~g,i }\Theta_{i, j_2 k2}+C_{p p~g,j_2 k_2 }\\
                &  + C_{p x~g,j_2\,m } \Theta_{m,k_2}\Big)+2\sum_{h=0}^P\Big(C_{x~g,m}\Theta_{m,k_2} C_{x~g,i}\Theta_{i,j_2} \\
&+C_{x~g,m}\Theta_{m,k_2}C_{p~g,j_2}+C_{p~g,k_2}C_{x~g,i}\Theta_{i,j_2}\\
&+C_{p~g,k_2}C_{p~g,j_2} \Big)
\\
 & \text{where}~~i,m=1,2,3 \dots N;~j_2,k_2=1,2,3 \dots M; \\
        & \text{and} ~g= 1,2,3, \dots S;
    \end{split}
\end{equation}
We now have presented a systematic approach for determining analytical gradient and Hessian to improve upon finite difference-based evaluation of gradients and Hessian which for some systems can serve as a difference between convergence and non-convergence of the system identification problem.

\section{Numerical example}

In this section, two widely used datasets, available for development and benchmarking in nonlinear system identification, will be used to test the presented general Newton method based optimization technique. Additionally, the performance of the proposed method will be compared for accuracy with the existing commercial system identification toolbox (Grey-Box Model Structure) available in MATLAB.

\subsection{Silver box model}
The first data set, which will be used, is the silver box model \cite{Pintelon.2005, Schoukens.2003}. The experimental data is available for download from \cite{unknown} and more information on the experimental silver box data can be found in \cite{Wigren.2013}. The silver box model describes an electronic implementation of a nonlinear system governed by the following nonlinear second-order differential equation.
\begin{equation}
\begin{split}
    &     m \dfrac{d^2 y(t)}{d t^2}+ d \dfrac{d y(t)}{d t}+k(y(t)) y(t) = u(t) \\
    & \text{where,} \quad k(y(t)) = a + b y^2 (t).
\end{split}\label{eqn25}
\end{equation}
The system consists of a moving mass $m$, a viscous damping $d$, and a
nonlinear spring $k(y(t))$. The sampling time
of the voltage signals’ measurements is $T_s = \dfrac{2^{14}}{10^7} s$ \cite{Wigren.2013,kocijan2018parameter}. Data-points from $10585$
to $11608$ for $u(t)$ and $y(t)$ are used as a training data set for estimation of the system parameters and initial conditions \cite{unknown}. Data points from $11609$ to $13655$ are used for validation of the model parameters. The cost minimization problem to identify  the unknown system parameters ($m$,$d$, and $k(y(t))$) and unknown initial condition $\dot{y}(t_0)$ can be posed as:
\begin{equation}
    \begin{split}
     &\underset{\displaystyle \boldsymbol{\overrightarrow{p}},\dot{y}(t_0)}{\text{min}} ~  J = \sum_{h=0}^P~\left(z_{{obs}}(t_h) -\tilde{z}(t_h) \right)^2 \\       
     & \text{such that}~~~  \dot{y}(t_h) = \dot{x}_1 (t_h) = x_2 (t_h) \\
       & \ddot{y}(t_h) =  \dot{x}_2 (t_h) = u(t_h)-\dfrac{d}{m} x_2 (t_h) - \dfrac{a}{m} x_1(t_h)- \dfrac{b}{m} x_1^3 (t_h) \\
       & \dot{\boldsymbol{\overrightarrow{p}}}  = \boldsymbol{0}; \quad  \\
        & \tilde{z} (t_h) =  C_x \boldsymbol{\overrightarrow{x}}(t_h);~~\text{where} ~C_x =\begin{bmatrix}
            1 & 0
        \end{bmatrix},\\
        &\boldsymbol{\overrightarrow{x}}(t_h)=\begin{bmatrix}
            x_1 (t_h) &
            x_2 (t_h)
        \end{bmatrix}^T; ~~\boldsymbol{\overrightarrow{p}}  = \begin{bmatrix}
            m &
            d &
            a &
            b 
        \end{bmatrix}^T
    \end{split}
\end{equation}
\noindent
The optimization problem is solved using the \textbf{fmincon} function of MATLAB, The implementation involves incorporation of both gradient and Hessian of the cost function, which are derived using the methodology introduced in this study, alongside the cost function. 
The initial guesses for unknown initial condition and system parameters are chosen to be $\dot{y}(t_0) = 0 $ and $m = 5.1025\times 10^{-6}$, $d=2.15\times 10^{-4}$, $a=0.968$, $b=3.976$ respectively (initial guesses in \cite{kocijan2018parameter} are used for reference to come up with the present choices of initial guesses). After going through $27$ iteration of optimization the final values for the unknown initial condition and system parameters are $\dot{y}(t_0) = 5.1112\times 10^{-9} $ and $m = 5.271\times 10^{-6}$, $d=2.1491\times 10^{-4}$, $a=0.9675$, $b=3.975$ respectively. It can be observed that the initial guesses are in a very close neighborhood of the final values for the parameters. This is because the relative order of the system parameters is quite high. 

\begin{figure}[htp]
\centering
{\includegraphics[width=0.485\textwidth]{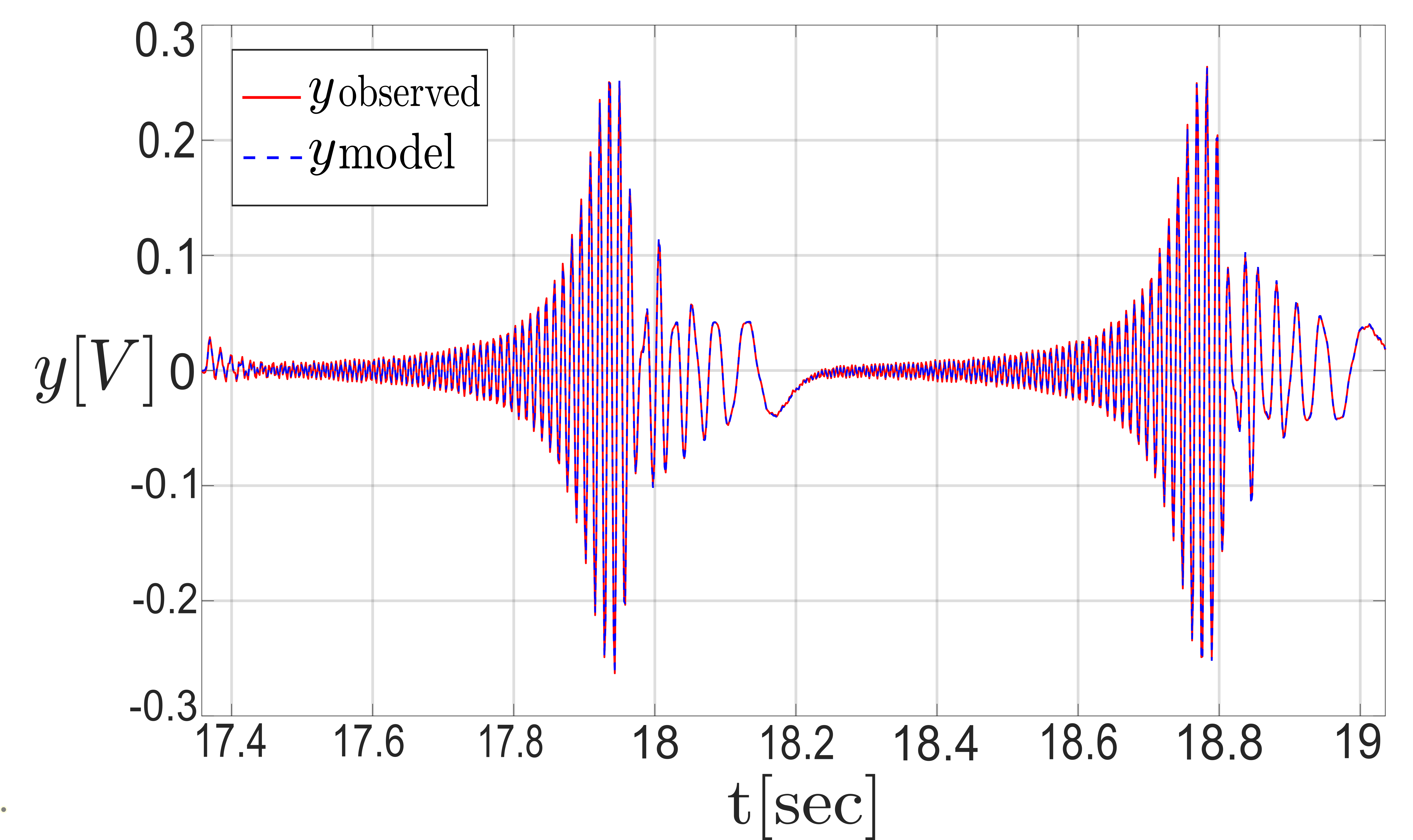}}%
\caption{Comparison of model $y$ data and observation $y$ data over the training data set.}
\label{Training data comparison}
\end{figure}
\noindent
Fig.(\ref{Training data comparison}) displays the comparison between the observed training data set and simulated model data derived from Eq.(\ref{eqn25}) with estimated system parameters and unknown initial condition.

\begin{figure}[htp]
\centering
{\includegraphics[width=0.485\textwidth]{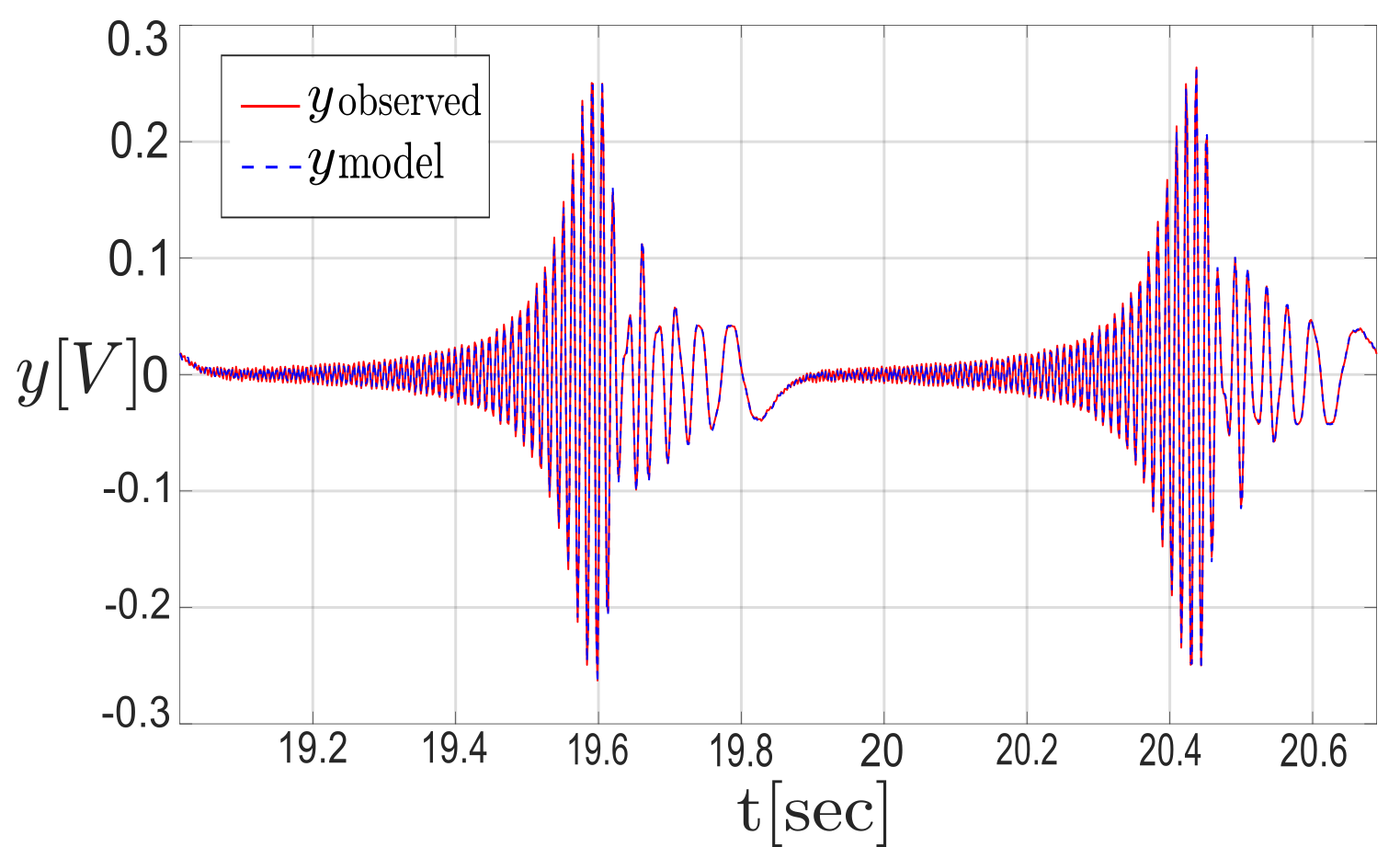}}%
\caption{Comparison of model $y$ data and observation $y$ data over the validation data set. }
\label{validation data comparison}
\end{figure}
\noindent
 Fig.(\ref{validation data comparison}) displays the comparison between the observed validation data set and simulated model data set derived from Eq.(\ref{eqn25}) with estimated system parameters albeit the initial condition, $\dot{y}(t_0)$ is unknown during the validation process. Hence, during validation process $\dot{y}(t_0)=0$ is assumed. In order to validate the model's simulation response across the entire validation signal, the goodness of fit (GOF) which is the complement of the normalized root mean square error (NRMSE) criterion is employed.
 \begin{equation}
     \mbox{GOF} = 1-\text{NRMSE} = 1 - \frac{||z_{obs} - \Tilde{z} ||}{||z_{obs} - E(z_{obs})||}
 \end{equation}
\noindent
where $z_{obs}$ is the vector of the validation data set. $\Tilde{z}$ is the vector of simulated model data set. $E(z_{obs})$ is the mean value of of $z_{obs}$. GOF attains a value of 1 in cases of a perfect match and it approaches $0$ when the variance of the model estimates is the same as the variance of the observed data. The obtained GOF value for the validation data set is $\text{GOF}=0.96258343=96.258343~ \%$ which can be considered as a very good fit. The same initial guesses were used when testing the performance of the system identification toolbox available in MATLAB. 

The MATLAB compiler aborts the computation because, for the chosen initial guesses for unknown initial condition and system parameters, the optimization could not be initiated when the relative tolerance and absolute tolerance of the cost function are set to the order of $10^{-14}$. Next, each initial guess was randomly perturbed within $20\%$ higher and lower than the original guess, and it was then used as an initial guess for optimization in the MATLAB toolbox. 
Out of 100 randomly chosen sets of initial guesses, not a single set within the vicinity of the original initial guesses could initiate the optimization process in the MATLAB compiler. Next, the relative tolerance and absolute tolerance of the cost function are set to the order of $10^{-12}$, and this time only three sets out of 100 randomly chosen sets of initial guesses within the vicinity of the original initial guesses could initiate the optimization process in the MATLAB compiler.
Same process is also followed while setting the relative tolerance and absolute tolerance of the cost function to the order of $10^{-10}$. This time, both the original choice for initial guesses and any random perturbation within its vicinity successfully initiated the MATLAB optimizer, allowing for the estimation of both unknown initial conditions and system parameters. In all cases for different order of relative tolerance and absolute tolerance of the cost function GOF of the model is measured using the obtained optimal unknown initial condition and system parameters. Table 1 compares the best goodness-of-fit (GOF) achieved for each case of relative tolerance (ReTol) and absolute tolerance (AbsTol) orders using both the MATLAB toolbox and the proposed system identification method. This allows for a clear comparison of performance between the two methods.

\begin{table}[h!]
\normalsize
    \centering
    \begin{tabular}{|c|c|c|c|}
    \hline
    \multirow{2}{*}{GOF}   &  \multicolumn{3}{c|}{~ReTol~,~AbsTol~}  \\ 
    \cline{2-4}
    \multicolumn{1}{|c|}{} & 
       $10^{-14}$  &  $10^{-12}$ &  $10^{-10}$\\
         \hline
         Proposed Method & $96.527\%$ & $96.258\%$ & $96.251\%$\\
         \hline
         MATLAB Greybox & Aborted & $55.627\%$ & $76.782\%$\\
         \hline
    \end{tabular}
    \caption{Proposed and Matlab Grey-Box system identification toolbox comparison}
    \label{tab:my_label}
\end{table}

\subsection{Two-tank problem}
The second data set is associated with the two-tank problem that has been used for testing system identification methods widely in the literature \cite{Wigren.2013}. The system dynamics is governed by the following nonlinear differential equation:
\begin{equation}
    \begin{split}
        & \dot{x}_1 (t) = - p_1 \sqrt{x_1 (t)} + p_2 u(t) \\
        & \dot{x}_2 (t) = -p_3 \sqrt{x_2 (t)} + p_4 \sqrt{x_1 (t)}
    \end{split}\label{eqn28}
\end{equation}
\noindent
where $x_1 (t)$ and $x_2 (t)$ represent the water level in tanks 1 and 2 respectively. The objective is to estimate the system parameters $p_1, p_2, p_3$ and $p_4$ following a gradient descent optimization technique, where the gradient and Hessian of the cost function are derived using the methodology introduced in this study. In this example, the complete data set ($501$ data points with a sampling period of $5.0~s$) will be used for system identification. The data set is available in \cite{unknown}. Next, using estimated system parameters water level in tank 1 ($x_1 (t)$) and tank 2 ($x_2 (t)$) will be derived from the model governing differential equation. Finally, $x_2$ from the model simulation and experimental observation will be compared. The cost minimization problem as it is described in \cite{Wigren.2013} can be formulated as:
\begin{equation}
    \begin{split}
        \underset{\displaystyle \boldsymbol{\overrightarrow{p}}}{\text{min}} ~  J = &\sum_{h=0}^P~\left(z_{{obs}}(t_h) -\tilde{z}(t_h) \right)^2 \\  \text{such that} ~ &
        \dot{x}_1 (t_h) = - p_1 \sqrt{x_1 (t_h)} + p_2 u(t_h) \\
       & \dot{x}_2 (t_h) = -p_3 \sqrt{x_2 (t_h)} + p_4 \sqrt{x_1 (t_h)} \\
        & \tilde{z} (t_h) =  C_x  \boldsymbol{\overrightarrow{x}}(t_h);\\
        & \dot{\boldsymbol{\overrightarrow{p}}} = \boldsymbol{0}~~
         \text{where}~ C_x =  \begin{bmatrix}
            0 & 1
        \end{bmatrix}, \\ \boldsymbol{\overrightarrow{x}}(t_h)= & \begin{bmatrix}
            x_1 (t_h) &
            x_2 (t_h)
        \end{bmatrix}^T;
        ~~\boldsymbol{\overrightarrow{p}} =  \begin{bmatrix}
            p_1 &
            p_2 &
            p_3 &
            p_4 
        \end{bmatrix}^T
    \end{split}
\end{equation}
\noindent
The optimization problem has been solved using the \textbf{fminunc} function of MATLAB. The gradient and Hessian of the cost function, which are derived using the methodology introduced in this study, are user-provided in the implementation of the optimization along with the cost function. The initial guesses chosen for unknown system parameters are $p_1 = 0.04$, $p_2 = 0.02$, $p_3 = 0.02$, and $p_4 = 0.04$ (initial guesses in \cite{Nandi.2018} are used for reference to come up with the present choices of initial guesses). After $100$ iterations the optimization converge the system parameters to  $p_1 = 0.0418$, $p_2 = 0.0235$, $p_3 = 0.0221$, and $p_4 = 0.0590$. 
\begin{figure}[htp]
\centering
{\includegraphics[width=0.45\textwidth]{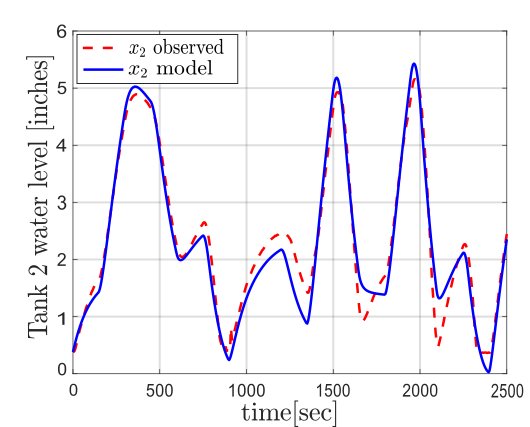}}%
\caption{Comparison of model $y$ data and observation $y$ data over the validation data set. }
\label{validation two tank problem}
\end{figure}
Fig.(\ref{validation two tank problem}) displays the comparison between observed training data set and simulated model data derived from Eq.(\ref{eqn28}) with estimated system parameters. The obtained GOF value for the training data set is $\text{GOF}=0.79088
=79.088~ \%$ which can be considered as a fairly good fit. In this example, for different orders of relative tolerance and absolute tolerance of the cost function, both the originally chosen initial guesses and any random perturbation within their vicinity successfully initiated the MATLAB optimizer, allowing for the estimation of both unknown initial conditions and system parameters. In all cases for different order of relative tolerance and absolute tolerance ($10^{-14} - 10^{-10}$) of the cost function, $\text{GOF}=0.79088
=79.088~ \%$ is achieved for the model.

\section{conclusion}
In this work, a generalized approach for determining the analytical gradient and Hessian of a cost function, constituted by state and parameter transition matrices, as well as higher-dimensional state and parameter transition tensors, is presented. Furthermore, it has been demonstrated that these matrices and tensors can be calculated by solving a set of ODEs. A general Newton method-based cost function minimization problem can be formulated, relying on gradient and Hessian information determined using the proposed approach. This method serves as a viable approach for identifying unknown system parameters and unknown initial conditions for the system states. When tested using two benchmark datasets, the proposed method for identifying system parameters and initial conditions has shown greater robustness against numerical instability, resulting in improved accuracy (goodness-of-fit) compared to the existing commercial system identification toolbox available in MATLAB.  

\bibliography{ifacconf}             
                                                   







\end{document}